\documentclass[twocolumn,showpacs,amsmath,amssymb,prd,nofootinbib]{revtex4}
\usepackage{epsfig}
\def\sss{\scriptscriptstyle}
\def\^#1{^{\sss #1}}
\def\_#1{_{\sss #1}}
\def\beq{\begin{equation}}
\def\eeqno#1{\label{#1}\end{equation}}
\def\kms{{\rm km~s^{-1}}}

\def\kpc{{\rm kpc}}
\def\mpc{{\rm Mpc}}

\def\msun{M\_{\odot}}
\def\az{a\_{0}}
\def\baz{\bar a\_{0}}

\def\l0{\ell\_{0}}

\def\l{\lambda}

\def\m{\mu}
\def\n{\nu}

\def\a{\alpha}

\def\RM{r\_M}

\def\MB{M\_B}
\def\MD{M\_D}
\def\gB{g\_B}
\def\gD{g\_D}
\begin{document}

\title{{\bf Perspective on MOND emergence from Verlinde's ``emergent gravity'' and its recent test by weak lensing}}
\author{Mordehai Milgrom}
\affiliation{Department of Particle Physics and Astrophysics, Weizmann Institute of Science, Rehovot 76100, Israel}

\author{Robert H. Sanders}
\affiliation{Kapteyn Astronomical Institute, Groningen, NL}

\begin{abstract}
We highlight phenomenological aspects of Verlinde's recent proposal to account for the mass anomalies in galactic systems without dark matter -- in particular in their relation to MOND.
Welcome addition to the MOND lore as it is, this approach has reproduced, so far, only a small fraction of MOND phenomenology. Like previous suggestions -- no more heuristic or less inspired -- for deducing MOND phenomenology from deeper, microscopic concepts, the present one is still rather tentative, both in its theoretical foundations and in its phenomenology.
What Verlinde has extracted from this approach, so far, is a formula -- of rather limited applicability, and with no road to generalization in sight -- for the effective gravitational field of a spherical, isolated, static baryonic system.
This formula cannot be used to calculate the gravitational field inside, or near, disk galaxies, with their rich MOND phenomenology. Notably, it cannot predict their rotation curves, except asymptotically. It does not apply to the few- or many-body problem. So, it cannot give, e.g., the two-body force for finite masses (such as two galaxies), or be used to conduct N-body calculations of galaxy formation, evolution, and interactions.
The formula cannot be applied to the internal dynamics of a system embedded in an external field, where MOND predicts important consequences. MOND is backed by full-fledged, Lagrangian theories that are routinely applied to all the above phenomena, and more. Verlinde's formula, as it now stands, strongly conflicts with solar-system and possibly earth-surface constraints, and cannot fully account for the mass anomalies in the cores of galaxy clusters (a standing conundrum in MOND).
The recent weak-lensing test of the formula is, in fact, testing
a cornerstone prediction of MOND, one that the formula does reproduce, and which has been tested before in the very same way.
\end{abstract}
%\pacs{04.50.Kd, 95.35.+d}

\maketitle

\section{Introduction}
MOND (originally standing for Modified Newtonian Dynamics) \cite{milgrom83} is a paradigm that contends to account for the mass anomalies in the Universe without invoking dark matter (some formulations of MOND can account also for `dark energy'). Recent reviews of the MOND paradigm can be found in Refs. \cite{fm12,milgrom14c}, and its history is described in Ref. \cite{sanders15}.
MOND introduces into gravitational dynamics a constant, $\az$, with the dimensions of acceleration. It approaches standard Newtonian/Einsteinian dynamics for accelerations much above $\az$. In the opposite, deep-MOND limit of accelerations much below $\az$, MOND becomes space-time scale invariant.
It was noted at the time of MOND's advent \cite{milgrom83} that $\az$, as determined from galaxy dynamics, is very near characteristic cosmological accelerations. With present-day values, we have
\beq \baz\equiv 2\pi \az\approx cH_0\approx c^2(\Lambda/3)^{1/2}, \eeqno{coinc}
where $H_0$ is the Hubble constant today, and $\Lambda$ the observed equivalent of a cosmological constant.
\par
There is, by now, a
considerable lore of possible theories that embody the above basic tenets of MOND, and hence make all the salient MOND predictions. These include full-fledged, non-relativistic, Lagrangian theories such as AQUAL (for aquadratic Lagrangian) \cite{bm84}, and Quasilinear MOND (QUMOND) \cite{milgrom10a}. They are nonrelativistic limits of relativistic MOND formulations, such as TeVeS \cite{bek04}, MOND versions of Einstein Aether theories \cite{zlosnik07}, bimetric MOND (BIMOND) \cite{milgrom09}, nonlocal, pure-metric theories \cite{deffayet11}, and massive-bigravity theories \cite{bh15}. MOND theories are reviewed in Ref. \cite{milgrom15}.
\par
As is largely the case with Newtonian dynamics and general relativity, one could remain satisfied  with such theories if they are self-consistent, allow us to calculate much of what we want, and agree with the measurements. But, as is the case with our standard theories of gravitational dynamics, it is tempting to derive such theories as effective ones, stemming from a (microscopic) theory at a deeper level.
\par
There is an even stronger impetus to try and do so in the case of MOND. Indeed, it has been stressed many times (Ref. \cite{milgrom15} and references therein) that MOND might emerge as such an effective theory. This assertion takes a particular hint from the `coincidences' of eq. (\ref{coinc}).
It is also hoped that such a more fundamental theory would account for the transition that MOND entails from standard to scale-invariant dynamics as we cross the MOND acceleration constant $\az$. (This transition is encoded in the so called ``extrapolation function''; see below.)
For example, it was shown in Ref. \cite{milgrom99} how the dynamics of an accelerated observer in a de Sitter universe -- such as ours is approximately -- may feel the de Sitter background such that
the de Sitter acceleration $c^2(\Lambda/3)^{1/2}$ appears in local systems (e.g., in galaxy dynamics) in the guise of MOND's $\az$ .
\par
Following Verlinde's original advocacy of the entropic basis of Newtonian gravity in a Minkowski Universe \cite{verlinde11}, several suggestions
of obtaining MOND from entropic/holographic arguments on the background of a de Sitter universe have been put forth \cite{pikhitsa10,kt10,ho10,lc11,kk11,klinkhamer12,pa12,pazy13,smolin16}.\footnote{There have also been a number of suggestions, along different lines, on how to obtain MOND as an emergent phenomenology, e,g., the membrane paradigm \cite{milgrom02}, breaking
of Lorentz invariance at low acceleration \cite{sanders11,blanchet11},  superfluid boson condensates \cite{khoury16}, and others.}
These all hark back to Ref. \cite{milgrom99} and the connection between MOND's $\az$ and $\Lambda$ it pointed out.  In fact, we pointed out privately to Verlinde in January 2010 that if he considered his emergent gravity in a de Sitter background, something resembling MOND can emerge, pointing specifically to Ref. to \cite{milgrom99}.
\par
The new work \cite{verlinde16} is the most recent in these attempts, and in our judgement it cannot obviously be considered to lie on firmer grounds than some of the previous suggestions.\footnote{Verlinde refers to his approach as ``emergent gravity''. It seems though that it is not all of gravity that might be emerging in his approach, only the part connected with the mass anomalies. The gravity contribution due to baryons is still taken as the standard, general-relativity one.}

Reference \cite{verlinde16} has aroused much interest recently.
Because confusion is rife as to what exactly emerges and how the theory is related to MOND, we feel that it is necessary to put the idea into perspective, and that is our aim here.
\par
The theoretical argumentations in Ref. \cite{verlinde16} concerning entanglement entropy are, by and large, opaque to us; so we only comment on them in general terms, based on what we think we do understand, and on discussion with others.
\par
It is quite clear that the idea is not yet based on some underlying, full-fledged, microscopic theory.
The principles from which Ref. \cite{verlinde16} starts, as they now stand, are not well-defined, quantitative theories. To boot, to obtain the result arrived at, various assumptions and ansatzes are made along the way -- assumptions that are not so clearly justified. For example, Ref. \cite{verlinde16} assumes that the entanglement entropy of de Sitter space is distributed over the entire volume of the Universe. Also, the line of arguments involves the Newtonian potential, $\Phi$, of the local mass, as it appears in the weak-field limit of the metric. But then [e.g., in eq. (7.38) there] for a spherical system, $\Phi$ is taken as $\Phi(r)=-G\MB(r)/r$, with $\MB(r)$ the baryonic mass inside $r$, ignoring, by fiat, the contribution to $\Phi$ of mass outside $r$. Using the standard expression for the potential would give different results.\footnote{One can also ask why take the $\Phi$ that appears in the metric (which is used e.g., for predicting gravitational lensing) as that produced by baryons only; why not take, selfconsistently, the total emergent potential, including the contribution of the phantom matter.}
\par
So clearly the aspects of local dynamics arrived at do not follow as inevitable, logical consequences from some well established principles or theories.
From this -- and from what we know of the history of the subject -- in our judgement, Ref. \cite{verlinde16} was intent on deriving specifically MOND phenomenology from the arguments, and so ride on its successes, rather than start from some compelling set of assumptions and follow the road they lay, wherever it leads.
\par
It seems to us that this assertion applies also to the exact ratio between the MOND acceleration and the Hubble constant that is gotten. There is enough freedom of choosing numerical factors along the way, and it seems that this freedom might have been used to arrive at a factor that will reproduce the previously known relation (\ref{coinc}).
\par
In the final analysis, even with this heuristic freedom the argumentations of Ref. \cite{verlinde16} could so far lead to the reproduction of {\it only a small fraction} of the MOND phenomenology. This comes in the form of a formula relating the gravitational acceleration produced by baryons to some integral over the `phantom' acceleration (attributed to dark matter) in spherical systems.
Extending this restricted formula to more general mass distributions seems to be a tall order.
\par
As an important example, the formula cannot be used to
treat disk-like systems, and hence to predict rotation curves of disc galaxies from the baryon distribution, which is a forte of MOND.
\par
We say all the above not to negate Verlinde's approach, but to stress that it still has a long way to go.
\par
Verlinde's formula does reproduce the two basic MOND predictions of (1) an effective logarithmic potential asymptotically far outside a confined mass, and (2) of the mass-asymptotic-speed relation, which relates the normalization of the logarithmic potential to the total baryonic mass (the basis for the baryonic Tully-Fisher relation).
\par
The recent galaxy-galaxy-weak-lensing analysis of Ref. \cite{brouwer17}, tests exactly these two cornerstone predictions of MOND, which are shared by the various approaches that lead to these basic MOND predictions. Moreover, this very test has been done before \cite{milgrom13} using a different data set.
The new analysis is thus a welcome, independent vindication of these MOND predictions, important in light of earlier claims of some tension, which were based on inferior data \cite{tian09}.

\section{What MOND aspects Verlinde's formula reproduces}

The line of arguments and assumptions in Ref. \cite{verlinde16} lead in the end to the extraction of an effective formula, very limited in scope, for the effective gravity of a spherical, isolated, static, nonrelativistic, baryonic mass distribution.
\par
Let $\MB(r)$ be the baryonic mass enclosed within radius $r$ of such a system, producing an acceleration $\gB(r)=G\MB(r)/r^2$. This is not emergent in the present treatment, but taken from standard dynamics. The added acceleration, $\gD(r)$, deduced in Ref. \cite{verlinde16}, which would be attributed to dark matter (or `phantom matter'), obeys the relation
\beq \frac{1}{3}\langle g^2\_D\rangle _r=r^{-3}\int_0^r g^2\_D(r)\hat r^2 d\hat r=\gB(r)\az, \eeqno{verlinde}
where, $\langle g^2\_D\rangle _r$ is the volume average of $g^2\_D$ within the sphere of radius $r$.
The phantom mass invoked to produce this acceleration is then $\MD(r)=G^{-1}r^2\gD(r)$.
(Confusingly, Ref. \cite{verlinde16} denotes the MOND acceleration (our $\az$) as $a\_M$, and reserves $\az$ for our $6\az$. We use the standard MOND notation in what follows.)
\par
Then, the total acceleration felt by a test particle is
\beq g(r)=\gB(r)+\gD(r).  \eeqno{totacc}
Outside a spherical mass, $\MB$, where $\gB(r)=\MB G/r^2$, eq. (\ref{verlinde}) gives $\gD(r)=[\az \gB(r)]^{1/2}$,
so the total acceleration is
\beq g(r)=[\az \gB(r)]^{1/2}+\gB(r). \eeqno{jj}
Equation (\ref{verlinde}) can be compared with the MOND relation for a spherical mass in modified gravity theories such as AQUAL or QUMOND:
\beq g(r)=\gB(r)\n[\gB(r)/\az], \eeqno{moni}
with the MOND interpolating function satisfying $\n(y\gg 1)\approx 1$, and $\n(y\ll 1)\approx y^{-1/2}$, which makes $g$ a unique function of $\gB$ -- called the mass-discrepancy-acceleration relation (MDAR) or acceleration-discrepancy relation (ADR).\footnote{Equation (\ref{moni}) is also exact for circular orbits in any axisymmetric field (such as for the rotation curves of disc galaxies) in `modified-inertia' formulations of MOND \cite{milgrom94}.} In contrast, by eqs. (\ref{verlinde})(\ref{totacc}) the relation between the baryonic and phantom accelerations is not local, and gives a functional relation between $g$ and $\gB$  only outside the mass, but not inside (see e.g., Ref. \cite{tejedor16}).
It reproduces MOND behavior when we are outside a spherical mass, or far enough outside any mass, such that this mass can be considered a point (hence spherical) mass. There, it takes the strict MOND form (\ref{moni}) with
\beq \n(y)=1+y^{-1/2}.  \eeqno{xx}
\par
In the deep-MOND regime, $\gB\ll\az$, eq. (\ref{jj}), inside and outside a spherical mass gives
\beq \langle g^2\_D\rangle^{1/2} _r\approx \sqrt{3}[\gB(r)\az]^{1/2}, \eeqno{v}
compared with the MOND expression
\beq g(r)\approx \gD(r)\approx [\gB(r)\az]^{1/2}. \eeqno{vi}
For systems in which $\gD$ declines with radius $\gD(r)<\langle g^2\_D\rangle^{1/2} _r$.
Outside the mass, equations (\ref{v}) and (\ref{vi}) are the same since there $\gD(r)=\langle g^2\_D\rangle^{1/2} _r/\sqrt{3}$.
\par
Because $\gB r^3$ is an increasing function of $r$, $\gD^2$ from eq. (\ref{verlinde}) is positive. However, it is possible for the resulting $\gD(r)$ to fall faster than $r^{-2}$, giving $\MD(r)$ that decreases with $r$, implying negative phantom densities at some radii.
A case in point is a system with sharp boundary, whose density drops quickly at the surface. Then, $\gD$ drops with radius discontinuously, or quickly, implying a thin shell of finite surface density of `phantom matter' {\it having a negative mass}. For example, for a homogeneous, baryonic sphere of mass $\MB$ and radius $R$, we have just inside $R$: $\MD(R^-)=2(R/\RM)\MB$, with $\RM\equiv (G\MB/\az)^{1/2}$ the MOND radius of $\MB$. Just outside $R$, we have $\MD(R^+)=(R/\RM)\MB$, implying a thin shell at $R$ of mass
\beq M\_S/\MB=-R/\RM=-(g_s/\az)^{-1/2}, \eeqno{shell}
where $g_s$ is the (baryonic) surface acceleration.\footnote{Under some circumstances, MOND may also predict a spherical shell of `phantom matter' around masses \cite{ms08}. But these are of a totally different nature: They have positive density, appear not at matter boundaries, but at the MOND radius of a concentrated mass, $\RM$, and their width is determined not by the matter distribution, but by the form of the MOND interpolating function.}
\par
Across a thin layer of baryonic matter, where $\gB$ makes a finite jump, $\gD$ becomes infinite (or very large).
For example, for a thin, hollow, spherical shell, both $\gB$ and $\gD$ vanish inside the shell, but $\gB$ is finite outside; so, by eq. (\ref{verlinde}), in the shell, $\langle \gD^2\rangle^{1/2}\propto\delta^{-1/2}$, where $\delta$ is the shell thickness.\footnote{The total phantom mass in the shall is finite but there are positive and negative regions each with very large mass, giving rise to a large value of $\langle \gD^2\rangle^{1/2}$.}

None of these discontinuities occur in the above mentioned MOND theories where $\gD$ is a function of $\gB$, not of its derivative. It is yet to be seen if such corollaries of eq. (\ref{verlinde}) have testable observational consequences.

\par
Equation (\ref{verlinde}) does not appear to lend itself to any visible generalization beyond spherical, isolated mass distributions. The argumentation leading to an expression for $\gD(r)$ lies heavily on the symmetry of the system, and this does not seem a mere technical difficulty. It seems to us that there are matter-of principle obstacles to treating aspherical systems with this approach. For example, the procedure at hand requires defining some volumes within which one integrates the squared phantom acceleration. For a spherical system with a preferred center, these volumes might plausibly be taken as concentric spheres, as done here. But this cannot be generalized to general mass distributions. Also, Verlinde's basic relation, involves scalar, volume integrals, for example of the square of the phantom (vector) acceleration field. It is hard to see how one would then be able to determine the three component the field from just a single scalar integral, unlike the spherical case, where the acceleration is always radial so it is enough to determine its magnitude.\footnote{A vestige of this conundrum, even in the spherical case, is that eq. (\ref{verlinde}) does not permit to determine the sign of the phantom acceleration. As we saw, this formula does result in negative phantom densities under some circumstances. So we cannot use positivity of the density to fix the sign of $\gD$.}

Be all the above as it may, even accepting eq. (\ref{verlinde}), it has to be realized that it accounts for only a small fraction of the MOND phenomenology (see, e.g., the reviews \cite{fm12,milgrom14c}, and Ref. \cite{milgrom14}).

\subsection{Some remaining MOND desiderata}
Reference \cite{verlinde16} emphasizes that the (nonrelativistic) formula deduced is only valid for spherical, static, non-dynamical systems. Of course, these are not limitations of MOND theories, such as AQUAL or QUMOND. These are full fledged formulation that can be used to calculate everything about galactic systems from the baryon distribution.
\par
We list below some important predictions and phenomena described by MOND phenomenology that cannot be addressed at present with Verlinde's approach.

Verlinde's formula cannot be applied to disc-like systems, hence it cannot make the all-important predictions of rotation curves of disc galaxies from the baryon distribution -- arguably the flagship of MOND phenomenology.
In particular, it is silent on predicted MOND relations such as the acceleration-discrepancy relation \cite{mcgaugh16b,milgrom16a}, or the central-surface-density relation \cite{lelli16,milgrom16}.
The MOND aspects of rotation curves that are captured by eq.(\ref{verlinde}) are asymptotic flatness of the rotation curves , and the mass-asymptotic-speed relation (baryonic Tully-Fisher relation).
\par
The dynamics of finite $N$-body system cannot be calculated. For example, the force between two finite masses -- such as two galaxies -- is not covered. This enters, for example, the analysis of the Milky-Way-Andromeda dynamical history \cite{zhao13}.
\par
Also, AQUAL and QUMOND are routinely used for MOND $N$-body calculations of galaxy formation, evolution, interaction
\cite{brada00,tiret08,llinares08,renaud16,thies16}.
None of this is thinkable with the restricted eq. (\ref{verlinde}).
\par
This formula can also not account for the internal dynamics of a system that is falling within the field of a mother system, such as dwarf satellite galaxies or globular clusters falling in the field of a mother galaxy, or a galaxy in the field of a galaxy cluster. In MOND this is described by the external-field effect (EFE), and is clearly seen in the data \cite{milgrom83,mm13,mcgaugh16a,caldwell16,hosein16}.
\par
AQUAL and QUMOND are non-relativistic limits of suggested relativistic MOND theories. Not so with Verlinde's formula.
\par
Like MOND generally, Verlinde's approach does not yet address properly the issue of the cosmological mass anomaly that calls for cosmological dark matter (e.g., Ref. \cite{sanders16} p. 122).

\subsection{Effects in the solar system and on earth}
To apply the more general eq. (\ref{verlinde}) to the motions of planets (as test particles) in the solar system, one should use the special case
eq. (\ref{jj}). This, as we saw, is the same as the MOND formula for spherical systems, with
$\n(y)$ of the form (\ref{xx}). The equivalent form of the MOND $\m(x)$ interpolating function\footnote{A kind of inverse of $\n(y)$: if $x=y\n(y)$, then $y=x\m(x)$.} is $\m(x)=[(x+1/4)^{1/2}-1/2]^2/x$. While these satisfy the two MOND asymptotic limits, they are, otherwise, known to be unacceptable due to solar-system constraints \cite{milgrom83,sj06}. Expression (\ref{jj}) implies a gravitational acceleration that approaches Newtonian values much too slow in the high-acceleration limit, and would produce departures from general relativity that are orders of magnitude larger than known solar-system limits. For example, it would strongly conflict with planetary perihelion advance: As shown in Refs. \cite{milgrom83,sj06} compatibility with these requires that for $x\gg 1$, $\m(x)-1$ vanishes faster than $x^{-\a}$, with $\a\gtrsim 2$; whereas eq. (\ref{jj}) implies $\a=1/2$.
For example, the relative correction to the solar Newtonian force on the earth this implies is $\sim (G\msun/r^2\_{\bigoplus}\az)^{-1/2}\approx 10^{-4}$  ($r\_{\bigoplus}$ is the astronomical unit), orders of magnitude larger than the general relativistic correction. This is true for all the known solar planets.
\par
Outside but near the earth, eq. (\ref{jj})
gives a relative departure from Newtonian gravity $\delta g/g\sim  (g/\az)^{-1/2}\approx3\times 10^{-6}$, given that on earth $g\sim 10^{11}\az$. At a larger radius, $R=\a R\_{\oplus}$, we have $\delta g/g\sim 3\times 10^{-6}\a$.
\par
Beyond this correction, due to the fictitious shell of phantom matter discussed above, one expects from eq. (\ref{shell}) that $g$ exhibit a down-jump from just inside to just outside the surface of the earth. With the mass of the shell being $M\_S\sim -M\_{\oplus}(R\_{\oplus}/\RM\^{\oplus})$ ($\RM\^{\oplus}$ is the MOND radius of the earth), we would have  $\delta g/g\sim  -(g/\az)^{-1/2}$. These effects are about 10 times larger than the variations in $g$ due to solar-lunar tides, and much larger than the general-relativistic correction to the Newtonian acceleration, which are known to enter e.g., in GPS-satellite orbits.
\par
One will then have to say that, for some yet unknown reason, the results of Ref. \cite{verlinde16} do not apply to the solar system. But Ref. \cite{verlinde16} is silent on this issue.

\subsection{Galaxy clusters}
Reference \cite{verlinde16} claims that eq. (\ref{verlinde}), unlike the MOND relation, might account for the mass discrepancies in the cores of galaxy clusters. The hope is based on the fact that eq. (\ref{verlinde}) gives an acceleration that can by about a factor of 2 larger than the MOND prediction at some radii. This, however, is not quite correct. We have typically (certainly in some clusters) $\langle g^2\_D\rangle ^{1/2}/\az\sim a few$ in these cores (within $\sim 100-200 \kpc$), and also $\langle \gD^2\rangle ^{1/2}/\gB\gtrsim 10$  \cite{sanders03,angus08}. So, clearly, eq. (\ref{verlinde}) still misses by a factor of a few. Verlinde's lapse may be traced to his thinking (as he quotes from various sources), that the remaining discrepancy in cluster cores, with MOND, is only a factor of $\sim 2-3$. The statement made in the literature \cite{sanders03,angus08}, and which Ref. \cite{verlinde16} quotes, refers, however, to the outskirts of clusters, at $\sim (1-2)\mpc$, where $\gD/\az<1$, and $\gD/\gB<10$, very different from typical values in the cores. At the outskirts, the order unity differences between eq. (\ref{verlinde}) and the MOND expression, can get an extra factor $\sim 2$, and might indeed alleviate the discrepancy, but not in cluster cores (see, e.g., Ref. \cite{ettori16}).

\section{The lensing test}
Reference \cite{brouwer17} has recently tested eq. (\ref{verlinde}) using galaxy-galaxy weak lensing, which, in essence, maps the many-galaxies-average gravitation potential at large radii. These radii are typically so large that the galaxies probed can be approximated by a point mass, hence eq. (\ref{verlinde}) can be used to a good approximation. Also, most of the data employed come from beyond the MOND radius of the galaxies probed
\beq \RM\equiv \left(\frac{\MB G}{\az}\right)^{1/2}=11 \left(\frac{\MB}{10^{11}\msun}\right)^{1/2}\kpc. \eeqno{viii}
(Reference \cite{brouwer17} studied galaxies in the range $2\times 10^{10}\msun-10^{11}\msun$.)
What is then tested is the deep-MOND limit of eq. (\ref{jj}),
\beq g(r)\approx [\gB(r)\az]^{1/2}=(\MB G \az)^{1/2}r^{-1}.  \eeqno{ix}
Such a gravitation field correspond to accumulative phantom mass
\beq \MD(r)=\MB \left(\frac{r}{\RM}\right). \eeqno{x}
These expressions coincide -- not fortuitously, as we said above -- with the MOND prediction eq. (\ref{vi}).
\par
What one tests in this way are both the $r$ dependence of $g$ predicted to behave as $r^{-1}$ -- tantamount to asymptotic flatness of rotation curves -- and the dependence on the central baryonic mass, $\MB$, both cornerstones of MOND phenomenology.
\par
This very prediction of MOND was already tested in this very way, first in Ref. \cite{tian09}, with much poorer data, then in Ref. \cite{milgrom13} using the much improved data and analysis of Ref. \cite{brimioulle13}. Equation (3) in Ref. \cite{milgrom13} coincides with eqs. (\ref{ix}-\ref{x}) above.
The analyses of Refs. \cite{milgrom13,brimioulle13} probes, in fact, galaxies down to masses of $\sim 5\times 10^{8}\msun$, much below the lowest masses of $\sim 2\times 10^{10}\msun$ in the analysis of Ref. \cite{brouwer17}. (Lower masses can, of course, be probed only to smaller radii with a given sensitivity.)

The analysis of Ref \cite{brouwer17} is thus a reconfirmation test of MOND as well as of all the suggestions that reproduce the above  two predictions of MOND, including Verlinde's.
\par
As stated above, Verlinde's approach is still silent on the so called MOND EFE. MOND, however says that eqs. (\ref{ix}-\ref{x}) cannot be valid to any radius, unless the central galaxy is totally isolated.
This effect implies (\cite{milgrom83}, and other references mentioned above) that if a system (e.g., a galaxy) of mass $\MB$ is embedded in an {\it extended} gravitational acceleration field of value $g\_{ex}$, then the internal dynamics should return to approximate Newtonian behavior roughly beyond the radius, $r\_{ex}$, where $g(r)=g\_{ex}$, or
\beq r\_{ex}=\left(\frac{\az}{g\_{ex}}\right)\RM=550\left(\frac{g\_{ex}}
{0.02\az}\right)^{-1}\left(\frac{\MB}{10^{11}\msun}\right)^{1/2}\kpc. \eeqno{xii}
Beyond $\sim r\_{ex}$, the internal dynamics is expected to be roughly Newtonian with an enhanced gravitational constant $G_{eff}=G(\az/g\_{ex})=50 G(g\_{ex}/0.02\az)^{-1}$.

Reference \cite{brouwer17} says that they tried to include relatively isolated lens galaxies.
We have no way of assessing exactly the degree of isolation. And, in any event, non-isolation in the sense that there is a galaxy nearby does not mean non-isolation as regards the EFE. What matters is an extended field that is roughly constant all around the central galaxy, in and beyond $r\_{ex}$. As has been argued before (e.g., in Ref. \cite{milgrom13}), at large enough radii the mean external field due to large scale structure should set some limit on $r\_{ex}$. This field can be estimated to be a few percent of $\az$ (a few hundred $\kms$ in a Hubble time). So MOND predicts the lensing signal to fall below that predicted by eq. (\ref{ix}), beyond $r\_{ex}$, which for $g\_{ex}=0.02\az$ is between $300\kpc$ and $550\kpc$. This effect might already have beeen seen in the data, but it is hard to tell with the somewhat large quoted errors.

\end{document}